\documentclass[3p,review]{elsarticle}

\usepackage{hyperref}

\journal{Computer Physics Communications}

\usepackage[utf8]{inputenc}
\usepackage[T1]{fontenc}
\usepackage{lmodern}

\usepackage{textcomp}

\usepackage{amsmath, amsfonts}

\usepackage{booktabs}
\usepackage{rotating}

\usepackage{xcolor}

\usepackage{graphicx}
\graphicspath{{img/}}

\usepackage{nicefrac}

\usepackage{siunitx}
\usepackage{subfig}

\usepackage{url}
\usepackage{ulem}
\usepackage{wrapfig}

\newcommand{\maxload}{\ensuremath{{l_\text{max}}}}

\newcommand{\PerformanceGain}{\eta}
\newcommand{\correct}[2]{#2}

\begin{document}

\begin{frontmatter}

\title{A Systematic Comparison of Runtime Load Balancing Algorithms for Massively Parallel Rigid Particle Dynamics}

\author[erlangen]{Sebastian Eibl\corref{mycorrespondingauthor}}
\cortext[mycorrespondingauthor]{Corresponding author}
\ead{sebastian.eibl@fau.de}
\author[erlangen,cerfacs]{Ulrich Rüde}

\address[erlangen]{Friedrich-Alexander Universit\"{a}t Erlangen-N\"{u}rnberg, Cauerstr.~11, 91058 Erlangen, Germany}
\address[cerfacs]{CERFACS, 42 Avenue Gaspard Coriolis, 31057 Toulouse, Cedex 01, France}

\begin{abstract}
As compute power increases with time, more involved and larger simulations become possible. However, it gets increasingly difficult to efficiently use the provided computational resources. Especially in particle-based simulations with a spatial domain partitioning large load imbalances can occur due to the simulation being dynamic. Then a static domain partitioning may not be suitable. This can deteriorate the overall runtime of the simulation significantly. Sophisticated load balancing strategies must be designed to alleviate this problem. In this paper we conduct a systematic evaluation of the performance of six different load balancing algorithms. Our tests cover a wide range of simulation sizes, and employ one of the largest supercomputers available. In particular we study the runtime and memory complexity of all components of the simulation carefully. When progressing to extreme scale simulations it is essential to identify bottlenecks and to predict the scaling behaviour. Scaling experiments are shown for up to over one million processes. The performance of each algorithm is analyzed with respect to the quality of the load balancing and its runtime costs. \correct{}{Additionally an applied test case is used to judge the applicability of the best algorithms in real world applications.} For all tests, the waLBerla multiphysics framework is employed.
\end{abstract}


\end{frontmatter}


\section{Introduction}
Rigid body dynamics are widely used for the simulation of rigid particles. Well known methods in this field are the Discrete Element Method (DEM)~\cite{Cundall1979} and methods based on non-smooth granular dynamics~\cite{Stewart2000,Preclik2015,Preclik2017a}. Together with the increasing compute power of today's supercomputers, scenarios comprising billions $(10^{10})$ of non-spherical particles and contacts are possible~\cite{Preclik2015,Eibl2018}. To achieve simulations this large, a carefully engineered software is needed with a focus on highly parallel algorithms. The typical parallelization strategy for particle simulations is to partition the simulation domain into subdomains and assign them to different processes~\cite{Beazley1995,Plimpton1995}. One important aspect of this initial domain partitioning is to achieve an equal workload for all cores. However, since the simulated system is dynamic, and the particles may migrate between subdomains, the workload might be shifted during the simulation. This leads to load imbalances, that can slow down the whole simulation. To overcome this problem, the domain partitioning must be adapted dynamically throughout the simulation and/or the subdomains must be reassigned to different processes. Many simulation frameworks have therefore adopted load balancing and results are published for simulations of various sizes.

\subsection{Related Work}
Compared to rigid body dynamics, molecular dynamics simulations differ in some aspects, however, the load balancing problem is closely related. Therefore we also consider methods proposed in the context of molecular dynamics here. A slightly dated but still relevant review of different methods suitable for load balancing can be found in~\cite{Hendrickson2000}. Owen et al.~\cite{Owen2000} use load balancing based on the ParMetis~\cite{Karypis1998a} graph partitioning library \correct{and show results for simulations with up to 6 cores.}{to balance their combined FEM-DEM simulation. They use two applied test cases namely a 2D bucket filling and a 3D hopper filling example. Measurements with up to 6 cores are presented.} Deng et al.~\cite{Deng2000} present a \correct{dynamic}{runtime} load balancing approach for molecular dynamics simulations which deforms the domain partitioning at runtime. The initial rectangular grid is optimized by moving the corners of all subdomains individually in space to adjust to the simulation. Good quality of the partitioning is reported \correct{}{for an artificial checkerboard scenario with no acting forces.} \correct{which}{The load balancing} improves the runtime performance but no claims about the global quality of the runtime improvement are made. Scaling results for up to \num{256} cores are shown by Begau and Sutmann~\cite{Begau2015} for an optimized version \correct{}{implemented in the community MD code \textit{IMD}. They also use two applied test cases - a nanopillar compression and surface coating - to evaluate their load balancing}. Another famous way for doing domain partitioning is by using Orthogonal Recursive Bisection (ORB) \cite{Warren1993} or Recursive Coordinate Bisection~\cite{Berger1987}. These methods recursively subdivide the simulation space by separating planes trying to keep the workload on both sides of the separating plane equal. Adaptions of these methods for particle simulations are reported~\cite{Fleissner2007,Fleissner2008,Shojaaee2012}. Fattebert et al.~\cite{Fattebert2012} present a detailed analysis of their load balancing approach based on a domain partitioning with Voronoi cells implemented in \textit{ddcMD}. A steepest decent algorithm is used to adapt the Voronoi cells dynamically. Good scaling results are shown up to \num{65536} MPI ranks \correct{}{for a MD simulation with uniform particle density}. A kd-tree based load balancing approach as implemented in the molecular dynamics software package \textit{ls1 mardyn}~\cite{Heinecke2015} is reported to show\correct{s}{} good performance up to \correct{1024}{2048} MPI ranks~\cite{Niethammer2014} \correct{}{for a planar interface scenario}. \textit{GROMACS}~\cite{Hess2008}, a well-known molecular dynamics framework uses cells to partition their domain. These cells are adapted due to time measurements carried out during the simulation. Unfortunately no more detailed measurement-based analysis of the load balancing algorithm is available in~\cite{Hess2008}. The \textit{LIGGGHTS} DEM framework~\cite{Berger2015} uses a Cartesian grid of subdomains and the \correct{dynamic}{runtime} load balancing is performed with a recursive multi-sectioning algorithm based on previous split locations and aggregated particle sums. \correct{}{A silo discharge and a mixing process is used to test the load balancing.} Measurements of the load imbalance with different MPI/OpenMP configurations with up to \num{128} threads are presented~\cite{Berger2015}. Building on this work, Cintra et al.~\cite{Cintra_2016,Cintra_2016a} demonstrate a hybrid OpenMP/MPI parallelization implemented in \textit{DEMOOP}. A RCB based algorithm is used for domain partitioning and various methods are applied for particle sorting and distribution onto threads. Scaling results are shown for up to 64 cores for different \correct{}{real world test setups (hopper discharge, landslide)}. However, the parallel efficiency \correct{most of the time remains below 50\%}{of the simulation code drops rapidly with the number of processes involved}. Markauskas and Kaceniauskas~\cite{Markauskas2015} present a comparison of the RCB implemented in the Zoltan library~\cite{Boman2012} and the multilevel k-way algorithm of the ParMetis library~\cite{Karypis1998a} with regard to \correct{dynamic}{runtime} load balancing \correct{}{in a hopper discharge setup}. Comparisons with up to \num{2048} cores are conducted using the \textit{DEMMMAT\_PAR} code. A summary of the largest simulations with load balancing as reported by various authors is collected in Tab.~\ref{tab:MaxProcessesComparison}.

\subsection{Contribution and Outline}
\begin{table}
\centering
\begin{tabular}{ l c r r }
  \toprule
  reference & cores & MPI ranks & load balancing algorithm \\
  \midrule
  Owen et al.~\cite{Owen2000} & 6 & & subdomain deformation  \\
  \textit{IMD}~\cite{Begau2015} & \num{256} & \num{256} & subdomain deformation \\
  Fleissner et al.~\cite{Fleissner2007,Fleissner2008} & 16 & & ORB based \\
  \textit{ls1 mardyn}~\cite{Niethammer2014} &  & \correct{1024}{2048} & based on kd-trees \\
  \textit{DEMOOP}~\cite{Cintra_2016,Cintra_2016a} & \num{64} & & RCB based  \\
  \textit{ddcMD}~\cite{Fattebert2012} & & \num{65536} & based on Voronoi cells \\
  \textit{DEMMMAT\_PAR}~\cite{Markauskas2015} & \num{2048} & & Zoltan \& ParMetis   \\
  \textit{LIGGGHTS}~\cite{Berger2015} & \num{128} & \num{128} & recursive multi-sectioning \\
  our contribution & \textbf{\num{262144}} & \textbf{\num{1048576}} & octree based \& various algorithms \\
  \bottomrule
\end{tabular}
\caption{Largest simulation reported with load balancing.}
\label{tab:MaxProcessesComparison}
\end{table}

\correct{
  Although load balancing is widely used, only few authors evaluate the performance they gained by load balancing in detail. In this paper we will use a systematic setup which enables us to give an a-priori estimates on the expected performance gained by using load balancing. We then evaluate different load balancing algorithms for rigid body dynamics simulations and compare the results with our expectations. Our goal is also to assess these algorithms for the usability in high performance computing. This includes problem sizes which were run on \num{262144} cores with \num{1048576} MPI ranks in parallel. To our knowledge this exceeds previously published results significantly, see Tab.~\ref{tab:MaxProcessesComparison}. Additionally we analyze the runtime and resource costs that have to be paid for these algorithms. All experiments were conducted on one of today's largest supercomputers -- the Juqueen supercomputer located in Jülich, Germany. Unfortunately this supercomputer was shut down in May 2018.
}{
Although load balancing is widely used, only few authors evaluate the performance they gained by load balancing in detail. Also a comparison between the different publications is difficult since the setups vary considerably and depend on many parameters. In the first part of our evaluation we will use a systematic simulation setup. This enables us to give an a-priori estimates on the expected performance gained by using load balancing and allows us to eliminate many potential influences on our measurements. It is also chosen such that already a few time steps are enough to yield representative data. This helps to reduce the run times and thus reduce cost especially when expensive supercomputers with large processor numbers are used. We evaluate different load balancing algorithms for rigid body dynamics simulations and compare the results with our expectations. Our goal is also to assess these algorithms for the usability in high performance computing. This includes problem sizes which were run on \num{262144} cores with \num{1048576} MPI ranks in parallel. To our knowledge this exceeds previously published results significantly, see Tab.~\ref{tab:MaxProcessesComparison}. Additionally we analyze the runtime and resource costs that have to be paid for these algorithms. All experiments were conducted on one of today's largest supercomputers -- the Juqueen supercomputer located in Jülich, Germany. Unfortunately this supercomputer was shut down in May 2018. In the second part we use a hopper discharge test case to also evaluate the load balancing algorithms under real world conditions. These experiments are run on the new Juwels supercomputer with fewer processes to save computation time since more time steps are needed for the evaluation.}

The remainder of this paper is structured as follows. In Sec.~\ref{sec:OurApproach} we describe our load balancing environment comprising the domain partitioning (Sec.~\ref{sec:DomainPartitioning}), the general load balancing pipeline (Sec.~\ref{sec:ARandLB}) and all load balancing algorithms selected for testing (Sec.~\ref{sec:LoadBalancingAlgorithms}). This is followed by a description of the computing environment in Sec.~\ref{sec:ComputeEnvironment}, the metrics we employ to compare the algorithms in Sec.~\ref{sec:Metrics} and our test setup in Sec.~\ref{sec:SimSetup}. Subsequently, the evaluation of the performance of the algorithms is presented in Sec.~\ref{sec:MediumProblem} and Sec.~\ref{sec:LargeProblem}. \correct{}{A comparison of the load balancing techniques in a hopper discharge scenario follows in Sec.~\ref{sec:HopperDischarge}}. Sec.~\ref{sec:Conclusion} summarizes the insight gained and discusses possible future lines of investigation.

\section{Our approach}
\label{sec:OurApproach}
Our approach targets millions of processes and beyond. The most important design criterion is therefore the strict locality of data and algorithms. Most data needed in a rigid particle dynamics simulation can be stored locally with only additional data from neighboring subdomains. This includes particle data, collision data, and all data needed for resolving collisions. However, the domain partitioning is typically stored on every process. This results in memory requirements that grow like $\mathcal{O}(p^2)$, with $p$ being the number of processes. A memory complexity like this limits the number of processes that can be used until all memory is depleted. Therefore the runtime and memory complexity is already an essential concern that must be addressed in the early stages of the algorithm and software design process. In the following sections we will outline our approach that leads to excellent scalability also for very large simulations~\cite{Schornbaum2016,Schornbaum2017,Eibl2018}.

\subsection{Domain Partitioning}
\label{sec:DomainPartitioning}
We use a distributed forest of octrees to organize our simulation domain. To generate the partitioning, we decompose the simulation domain into a grid of equally shaped brick-like domains. Each of these bricks acts as the root of an octree. Depending on the setup we then start to refine each octree individually by subdividing the corresponding subdomain at its center into eight equal subdomains. Each of these subdomains forms a branch of the octree. The procedure is then repeated for every branch. The subdomains used for the simulation are the leaves of the octree. We also impose additional constraints on our data structure to make it possible to handle and refine the domain efficiently during a simulation. A parent node is always split exactly at its center and neighboring subdomains can differ by at most on level of refinement. This helps to keep the number of neighbors of each subdomain bounded. \correct{}{These restrictions reduce the flexibility of the domain partitioning to cover the simulation area optimally. However, the benefits of this data structure are that it can be stored in a distributed fashion and it is easy to determine the neighborhood of subdomains. Both features are essential for extreme scale simulations.} A more detailed description of this data structure and its distributed implementation can be found in~\cite{Schornbaum2016}. Building upon this, the load balancing pipeline is described in the next section.

\subsection{Load Balancing Pipeline}
\label{sec:ARandLB}
The actual load balancing pipeline is a three step process. First weights are calculated for every subdomain. Two weights are needed for the subsequent load balancing process. One is the \emph{computational weight} which quantifies the work that must be performed to advance all particles within a subdomain by one time step. Typically this number is closely related to the number of collisions that need to be resolved in that subdomain. The time needed for collision detection and collision resolution scales essentially with the number of contacts. A second weight is the \emph{communication weight} which describes how much communication is involved in keeping the subdomain in sync with its neighbors. The exact computation of the weights in our setup will be described in Sec.~\ref{sec:SimSetup}.

When all weights have been determined, we refine the octree in areas with a high workload and coarsen it in low workload areas. This is done with the goal of maintaining a compromise. On the one hand we must keep the number of subdomains small and on the other hand generate small enough workload portions so that we can efficiently balance the workload. Since only whole subdomains can be moved between processes, the granularity of the workload limits the achievable workload balance. Note that the octree is not refined/coarsened for physical reasons as the simulation accuracy is not related to the subdomain size. 

In the final step, the leaf nodes (subdomains) are distributed among the available processes. There different algorithms can be chosen. The goal of these algorithms is to make sure that the workload of every process is similar. In the following we will briefly describe the algorithms used for this evaluation.

\subsection{Load Balancing Algorithms}
\label{sec:LoadBalancingAlgorithms}
In this paper we compare several load balancing algorithms for their suitability in large scale parallel particle simulations. The first class of algorithms is based on space filling curves (SFC). SFCs map the 3D space onto a 1D curves~\cite{Bader2012}. This is used to construct a linear ordering of all subdomains. The load balancing itself then searches for cuts within this linear ordering to generate $p$ equal partitions. The partitioning must be computed with respect to the aggregated load of each partition, defined as the sum of the loads of each octree leaf in it. The partitions are then distributed among the $p$ processes. Morton~\cite{Morton1966} and Hilbert Peano~\cite{Campbell2003} space filling curves are used in this article. A property of these SFCs is the conservation of spatial locality. This is especially important in communication sensitive applications, like rigid body dynamics, since it keeps the communication distances small. But this advantage comes at a cost. SFCs inherently need global information about all weights and subdomains. To maintain consistency across all processes, all processes have to agree on the same subdomain distribution. This can only be achieved by broadcasting all weights to all processes. Like for the non distributed domain partition, this will eventually lead to a memory problem, since the memory needed here grows like $\mathcal{O}(p^2)$.

The second class of algorithms is based on load diffusion. Here load balancing is achieved by an iterative algorithm which tries to smoothen out load imbalances. The algorithm computes load gradients between processes and transfers subdomains from highly loaded processes to processes with less load~\cite{Cybenko1989}. In principle, the number of iterations has to be adjusted together with the number of subdomains to achieve a consistent good load balance. However, if the algorithm is run regularly, very large load imbalances should not occur and a fixed low number of iterations should be sufficient. This class of algorithms has shown promising results in very large scale fluid dynamics simulations~\cite{Schornbaum2017}. The major benefit of this algorithm is its strict locality which makes it a good candidate for extremely parallel simulations. However, due to its strict locality and since it is an iterative method, the quality of the load balance might be worse than for other algorithms. In particular, there is no guarantee that domains are kept together which might significantly hurt performance in communication sensitive applications.

The last class are graph based algorithms, as provided by the ParMetis~\cite{Karypis1998a} graph partitioning framework. In particular we use the Kway algorithm which is based on a multilevel k-way partitioning algorithm~\cite{Karypis1997,Schloegel2000}. The Geom\_Kway algorithm uses a space filling curve to compute the initial partitioning and then applies the Kway algorithm. Finally we also compare with Adaptive\_Repart which is a Unified Repartitioning Algorithm that combines remapping and diffusion-based repartitioning schemes~\cite{Schloegel2000a}.

\section{Performance Results}
In this chapter we will introduce the software framework and computer used for the performance evaluation. We then describe the metrics used for our comparison as well as the simulation setup. Following this introduction we present a detailed analysis of the performance data we gathered. 

\subsection{Framework and Supercomputer}
\label{sec:ComputeEnvironment}
All algorithms introduced in chapter \ref{sec:LoadBalancingAlgorithms} are implemented in the waLBerla multiphysics framework that is freely available at \url{www.walberla.net} and licensed under GPLv3. Previous results obtained with this framework show very good scalability for rigid body dynamics simulations~\cite{Preclik2015,Eibl2018}. All tests are executed on the Juqueen supercomputer located at Jülich Supercomputing Center (JSC) and ranked 22 in the TOP500 list\footnote{https://www.top500.org/} as of November 2017. This BlueGene/Q system has \num{28672} compute nodes~\cite{Gilge2014,Wautelet2014}. Each node is equipped with a single IBM PowerPC A2 processor that has 18 cores clocked at \SI{1.6}{GHz}, but only 16 cores are available for computing. Each processor supports 4-way simultaneous multi threading (SMT). The available memory for each node is \SI{16}{GiB} of RAM. The interconnect fabric is a 5D torus topology which features a bandwidth of \SI{16}{Gbit\per s} per link and direction~\cite{Chen2012}. This computer reached its end of lifetime in May 2018.

\subsection{Metrics}
\label{sec:Metrics}
Since the performance of a parallel code is always limited by the slowest process our first metric targets exactly this process. We define the \emph{max load per process} by
\begin{equation*}
l_\text{max} = \max_p \left\{ l_p \right\}
\end{equation*}
with $l_p$ being the accumulated load of all subdomains located on process $p$. As the load is used as an input parameter for all load balancing algorithms, this quantity should be close to the average load if the load balancing algorithm is working correctly.

With the second metric we want to measure the performance gained by using load balancing. The \emph{performance gain} is defined by
\begin{equation*}
\PerformanceGain = \frac{t_\text{before load balancing}}{t_\text{after load balancing}}
\end{equation*}
with $t$ being the time needed to complete one time step. To reduce the sensitivity to fluctuations we always take the average over 100 time steps. It is also important to make sure that the setup does not change drastically within the measurement period to exclude unwanted variability for this metric. We will take care of this with a specifically designed setup which will be described in Sec.~\ref{sec:SimSetup}.

Finally we also examine at the time needed for the actual load balancing pipeline $t_{lbp}$. This includes the time needed to refine/coarsen the domain, the balancing process and the migration of all subdomains. For the load balancing to be usable, this should be in the order of at most a few regular time steps. Otherwise the load balancing will dominate the overall runtime of the simulation. 

\subsection{Simulation Setup}
\label{sec:SimSetup}

\begin{figure}[h]
  \centering
  \includegraphics[width=0.5\textwidth]{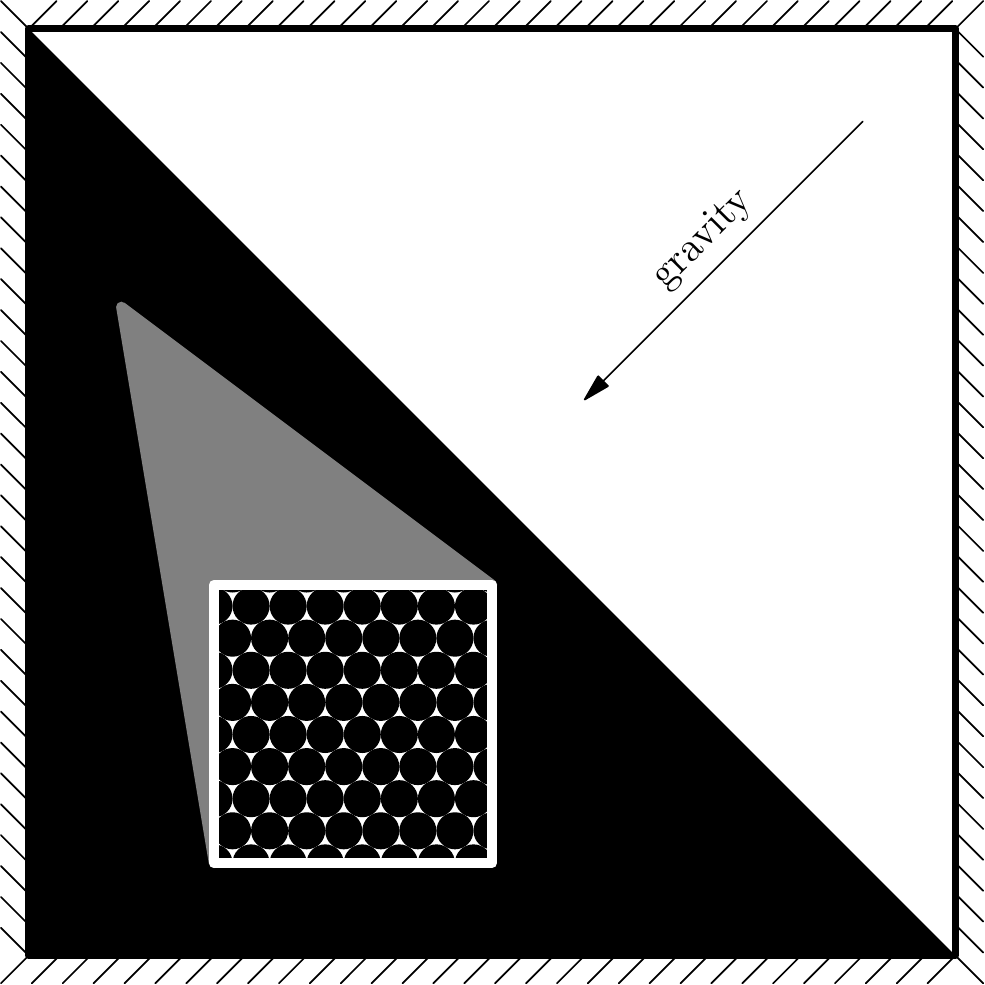}
  \caption{2d cross section along the $xy$ plane of the simulation setup. This setup is used to measure the performance of different load balancing algorithms. The simulation domain is confined by solid walls. The gravity is acting towards the lower left edge. All particles are arranged along that edge in a stable hexagonal close packing lattice. All particles are in contact with their neighbors. This assures that the setup will not change when the simulation is integrated in time. The simulation setup extends uniformly in the $z$ direction.}
  \label{fig:SetupVisualization}
\end{figure}

As the benchmark scenario, a particle configuration is chosen that does not move in time. This seems to be in contrast to why we need \correct{dynamic}{runtime} load balancing at all. However, we choose this setup deliberately to obtain a quantitative evaluation of the performance of the load balancing algorithm. It allows us to compare the runtimes before and after load balancing without the influence of a specific particle configuration. The setup itself is kept simple to enable us to give an a-priori estimate for the maximum performance gain achievable. With this setup we deliberately eliminate many factors which would be difficult to control and which could otherwise lead to biases in our measurements. \correct{}{Besides that, it enables us to retrieve meaningful results within few time steps. This saves precious computation time especially in large scale simulation.}

The general simulation domain is a box confined by solid walls. The box is filled to a certain height with a large number of spherical particles. Since the gravity points from the center of the box towards one edge, parallel to the $z$ axis, we start the filling process at this edge. The spherical particles are arranged in a hexagonal close packing (hcp) lattice. All particles are in contact with their neighboring particles. This makes sure that this particle configuration is already optimal and therefore the particles stay at rest. In this configuration the arrangement of particles and domain partitions is uniform in the $z$-direction so that we deal essentially with a two dimensional situation. The simulation size can be adapted by growing the domain along the $z$ axis without changing the properties of the setup. A visualization of the setup with a half filled box is shown in Fig.~\ref{fig:SetupVisualization}. For the time integration of the system a non-smooth granular dynamics algorithm is used~\cite{Preclik2015}.

\begin{figure}[h]
  \centering
  \subfloat[before]{\includegraphics[width = 0.45\textwidth]{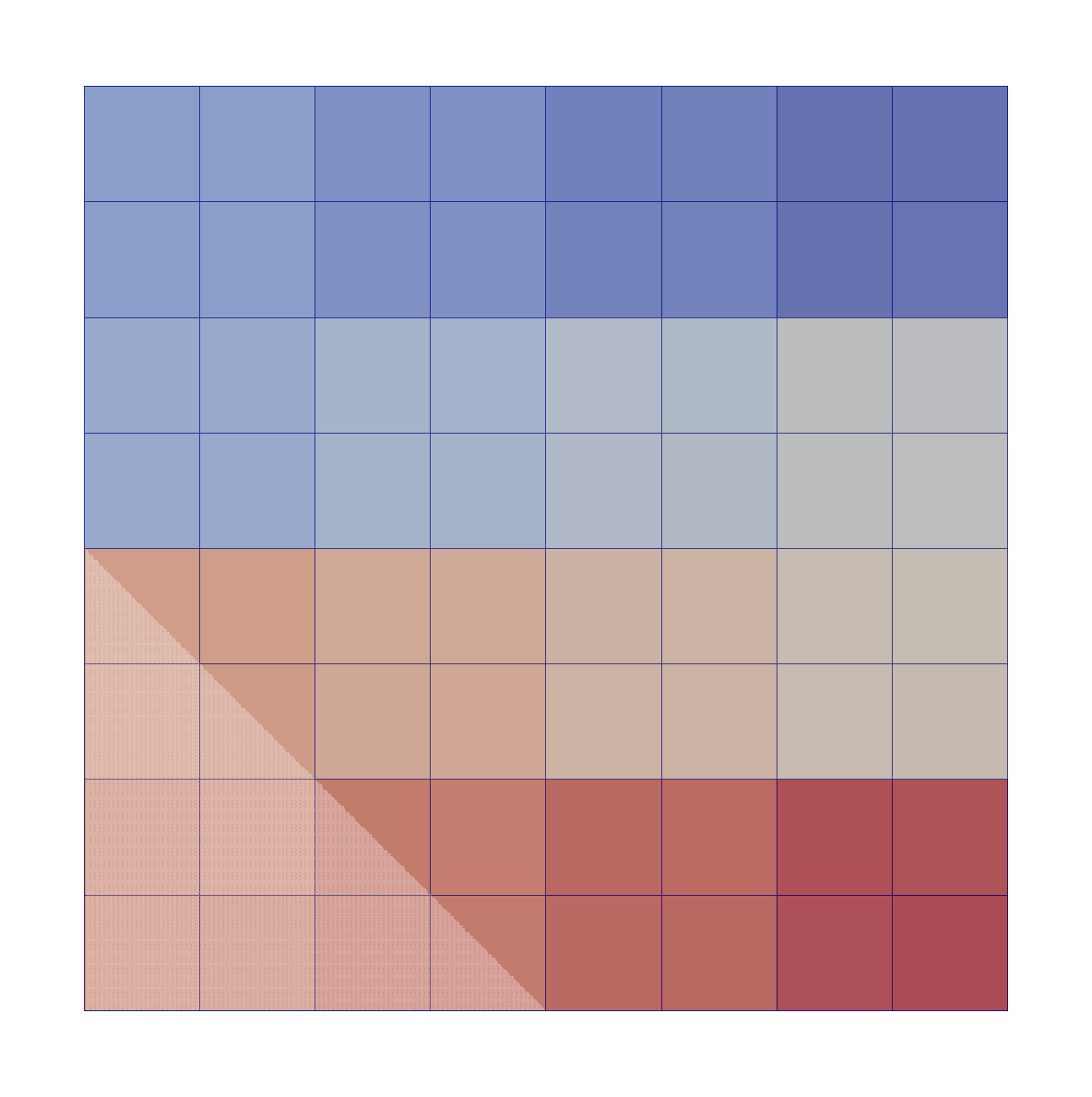}}
  \hfill
  \subfloat[after]{\includegraphics[width = 0.45\textwidth]{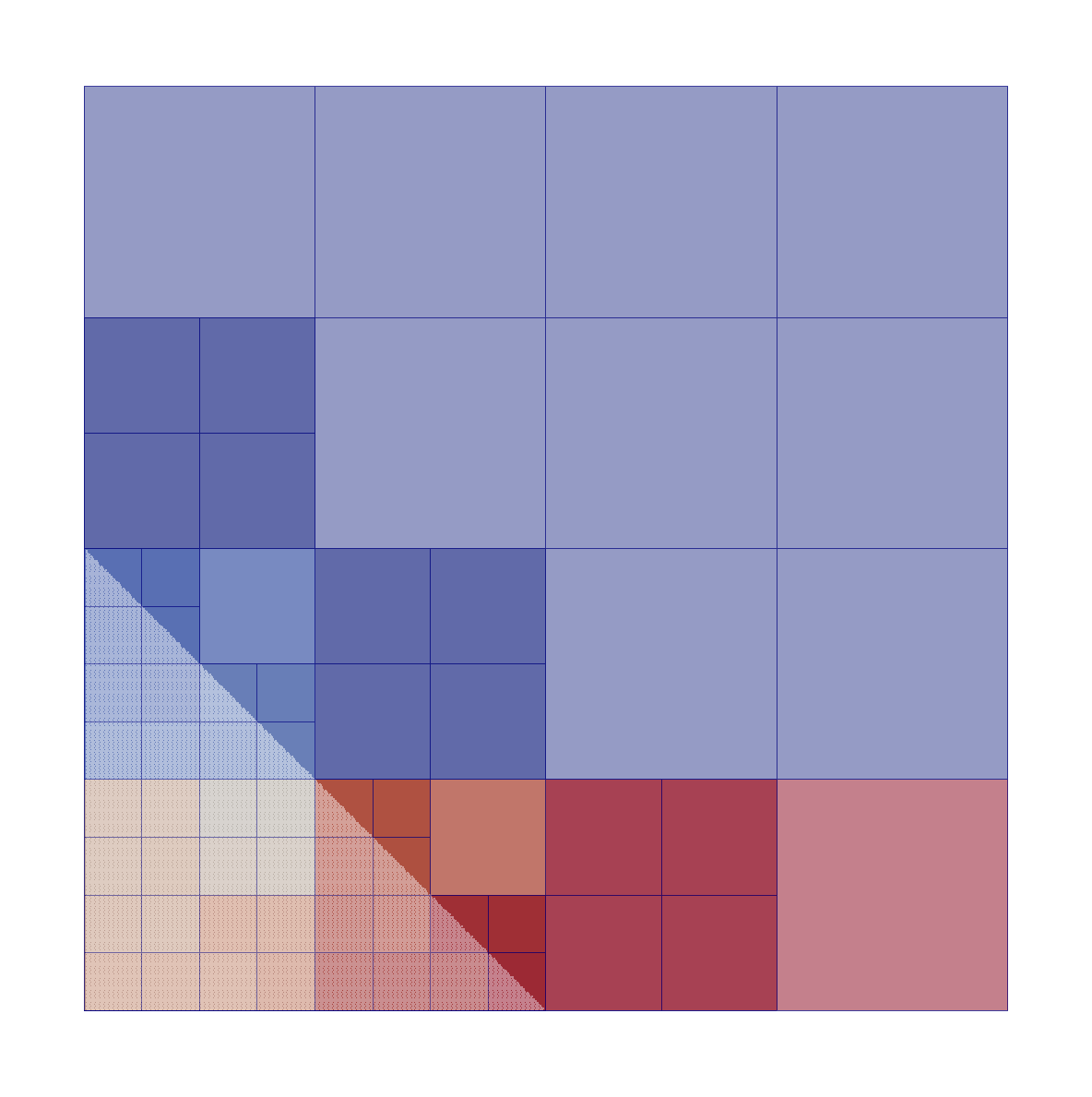}}
  \caption{Domain partitioning before and after the load balancing pipeline. Again a 2d cross section along the $xy$ plane of the simulation setup is shown. The simulation domain is filled by $\nicefrac{1}{8}$ with spherical particles. The particles are located at the lower left edge as indicated by the shading in lighter color. All subdomains are colored to represent the process they belong to. The initial partitioning uses 128 processes and 128 subdomains such that a 1:1 mapping is possible. The grid partitioning of the simulation domain is $4\times4\times1$. Before the load balancing every octree is refined exactly once ( see subfigure a) ). During the refinement/coarsening process of the load balancing pipeline empty subdomains in the top right corner get merged and subdomains in the lower left get split. Note also the medium sized subdomains in the middle which are needed to fulfill the refinement constraint. Neighboring subdomains are allowed differ by at most one level of refinement. For subfigure b), a Hilbert space filling curve is used to calculate the distribution of the subdomains to the processes.}
  \label{fig:DomainPartitioning}
\end{figure}

This scenario permits no efficient domain partitioning if a balanced octree is used. This can bee seen in Fig.~\ref{fig:DomainPartitioning}a). The initial domain partitioning has exactly as many subdomains as processes available. This suboptimal initial configuration is chosen deliberately. In real world applications chances are that the user cannot choose a perfect domain partitioning for the setup due to restrictions by the framework or simply because the perfect partitioning is not known. It is also possible - and gets more likely throughout the simulation - that the particle configuration changes such that no static domain partitioning is appropriate. Therefore, the load balancing pipeline must find the best possible partitioning regardless of initial user input and at runtime. In our experiment, the initial setup is advanced by 100 time steps and the runtime is measured. This measurement acts as a baseline to judge the speedup that can be achieved by the load balancing pipeline.

The first step of the load balancing pipeline is the weight assignment for each subdomain. This is independent of the load balancing algorithm. Since the particles are arranged on a hcp lattice, we assume a contact number of 12 and a direct relation between the number of particles and the contacts that have to be resolved. Therefore we use the number of particles per subdomain as its computational weight. For the ParMetis based algorithms we use the area of the interface between two adjacent subdomains as an estimate for the communication weight. The next step is also independent of the load balancing algorithm used. The octree gets refined/coarsened to achieve a better granularity for the subsequent load balancing step. The refined domain partitioning is shown in Fig.~\ref{fig:DomainPartitioning}b). A simple method based on thresholds on the number of particles per subdomain is used to decide whether a refinement must be performed. After the refinement the load balancing algorithm is executed. Due to the special considerations taken into account while designing the test scenario, the diffusive load balancing is guaranteed to converge in all simulations within a fixed number of iterations. We therefore keep the number of iterations constant throughout all simulations. With the new domain partition the simulation is then advanced again for 100 time steps, taking time measurements. 

To assess the suitability for highly parallel applications, we conduct a weak scaling~\cite{Hager2010} study, i.e. we increase the computational effort by the same factor by which we increase the computational power. Therefore we extend the original simulation domain in the $z$ direction and at the same time we increase the number of cores used. The domain partition in the $xy$ plane has $8\times8$ cubic subdomains. Results for two scenarios with a different initial imbalance are shown in Sec.~\ref{sec:MediumProblem} and Sec.~\ref{sec:LargeProblem}.

\subsection{Medium Size Problem}
\label{sec:MediumProblem}

\begin{figure}[h]
  \centering
  \subfloat[maximum load per process]{\includegraphics[width = 0.45\textwidth]{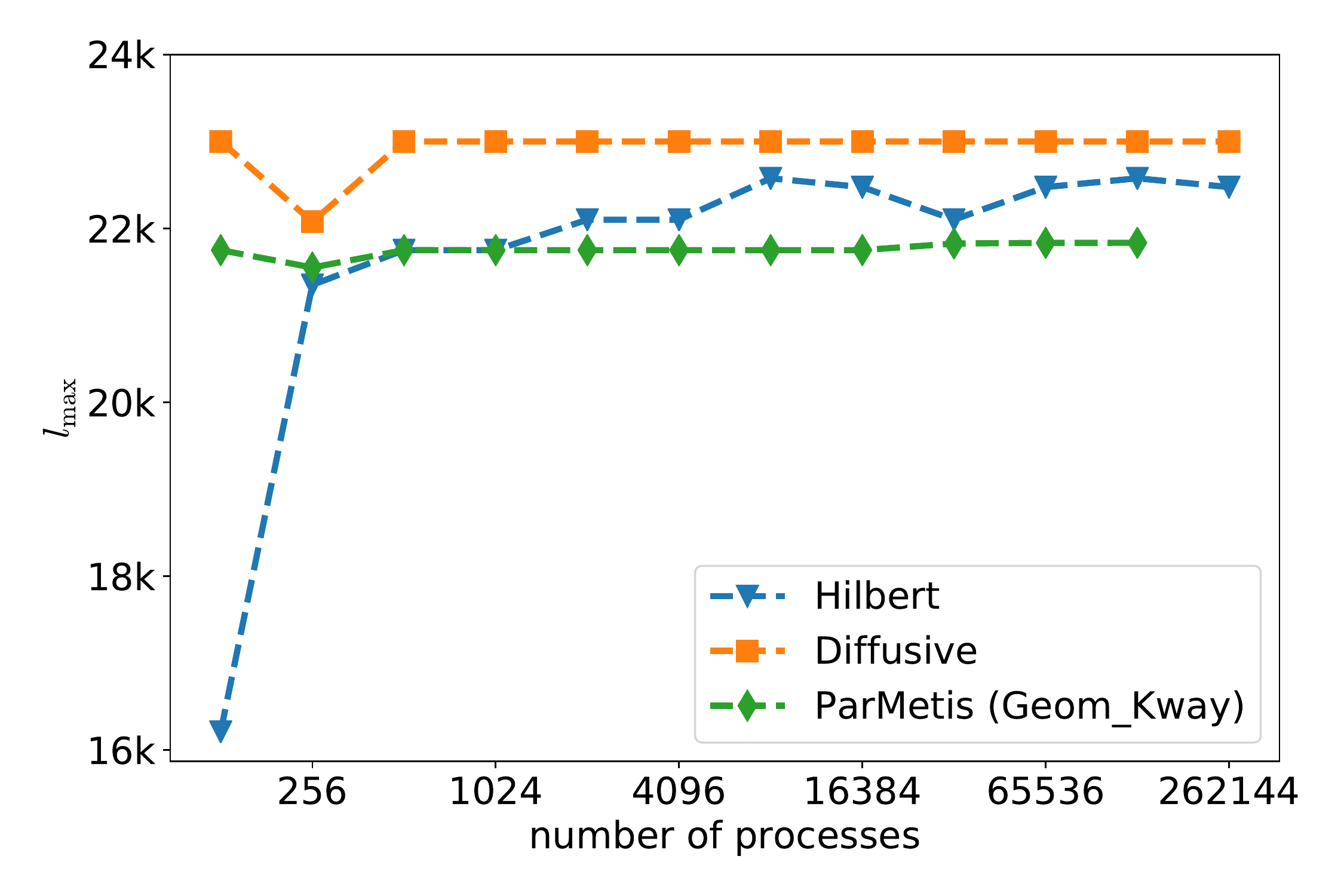}}
  \hfill
  \subfloat[performance gain]{\includegraphics[width = 0.45\textwidth]{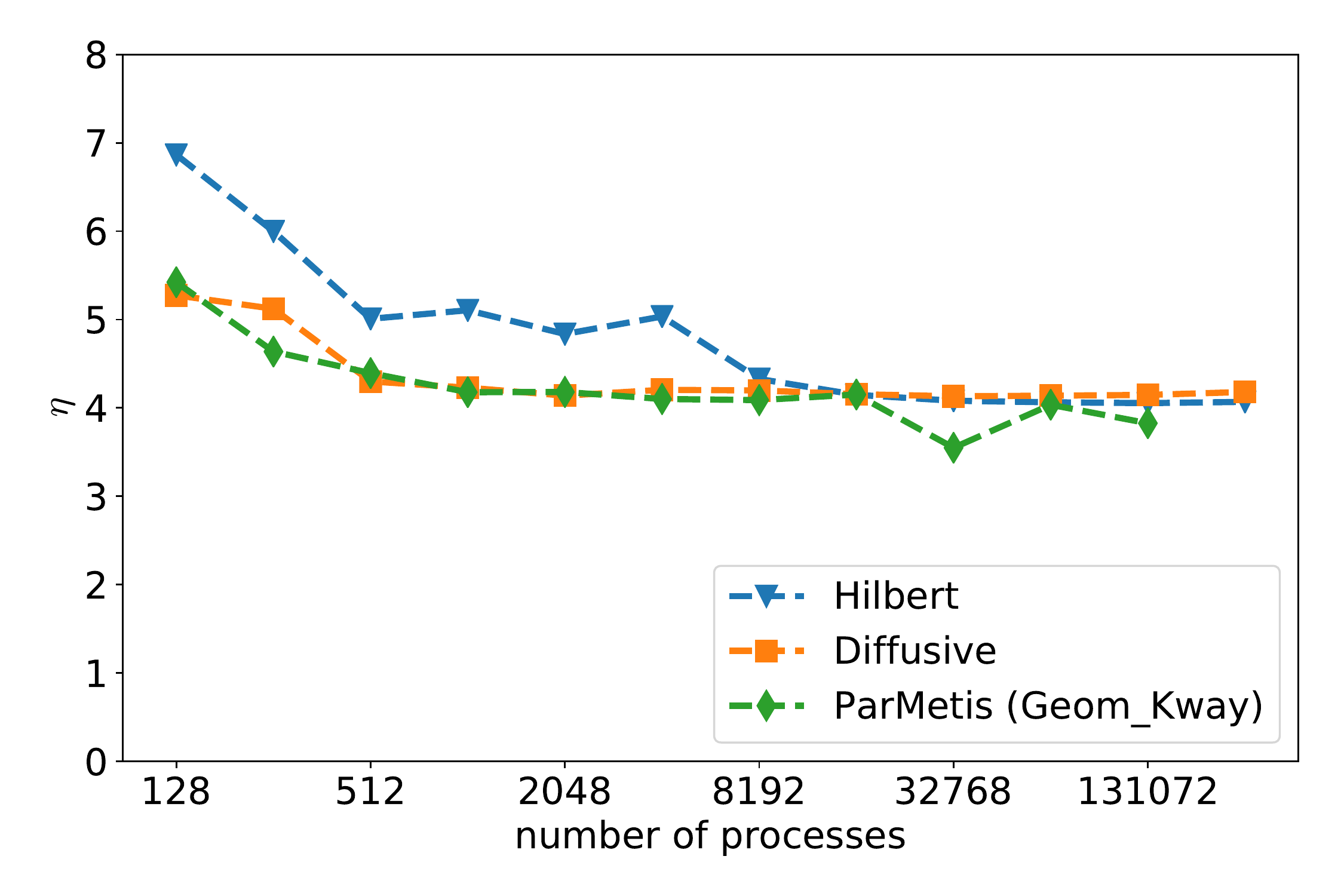}}
  \caption{On the left side the quality of the load balancing pipeline is shown. The right side shows a comparison of three different load balancing algorithms. The performance is given relative to the performance before the load balancing. All data is evaluated for different simulation sizes.}
  \label{fig:HighLoad}
\end{figure}

The first evaluation uses a simulation box, of which $\nicefrac{1}{8}$ is filled with particles. This means that at the start only $\approx\nicefrac{1}{8}$ of the processes are associated with subdomains that contain particles. Empty subdomains impose almost no workload so the whole work is done by $\approx\nicefrac{1}{8}$ of the processes. The size of the particles is chosen such that almost \num{90000} particles are located in each subdomain when it is completely filled. To cope with the limited amount of memory available per node, the SMT feature is not used, i.e. only \num{1} process is spawned per core.

Every time step of the simulation consists of computational work and communication. In theory, by using all available processes efficiently, the maximum load per process can be reduced by a factor of 8 which should lead to a performance gain of 8. However, due to the refinement process, the total communication weight gets increased by a factor of 2 (eight times as many subdomains with a quarter of the original surface area). Together with also an increase of 8 in the available network resources, this should lead to a performance gain of 4 for the communication part. Since the time spent in the communication also depends on the mapping of the processes onto the machine the actual performance can vary in both directions. However, a good load balancing algorithm that respects spatial locality can optimize the mapping by placing adjacent subdomains on neighboring ranks. The overall performance gain depends on the relation between the time spent in computation and communication which is not clearly identifiable since communication is interleaved with computations in our implementation.

We evaluate the performance of the Hilbert space filling curve, the diffusive algorithm and the Geom\_Kway algorithm of the ParMetis library for this scenario. We start by analyzing the load distribution after the load balancing. Note that in an ideal case with perfect load distribution the number of particles per process would be approximately $\num{11000}$. Note also that the granularity of the load balancing is $\num{90000} / 8 \approx \num{11000}$ - all subdomains which are completely filled with particles are refined into 8 smaller subdomains. Since only whole subdomains can be transfered between processes this limits the accuracy of the load balancing. Reaching the optimal load balance is almost impossible as one misplaced block changes the load by \num{11000}. With this in mind all tested load balancing algorithms achieve a very good load balance for all setup sizes only overloading the slowest process by one subdomain. Fig.~\ref{fig:HighLoad}a) displays the maximum load per process after load balancing. We also do not observe differences in the load balancing optimality achieved by the algorithms. With this information we can refine our estimate on the performance gain of the computational part. The load for highly loaded processes is reduced by $\frac{\num{90000}}{\num{22000}} \approx \num{4.1}$. This performance gain is now equivalent to the performance gain of the communication. We therefore assume that the total performance gain is 4.

The performance gain is displayed in Fig.~\ref{fig:HighLoad}b). The best performing algorithm is the Hilbert space filling curve that reaches a performance gain of almost 7 for \num{128} processes. With increasing domain size the performance gain for all algorithms worsens and converges to 4 as expected. One possible reason for deviating performance in the beginning is the fundamental change in the simulation setup. For higher process numbers a larger simulation domain is needed. This is achieved by growing the simulation domain along the $z$ axis introducing more subdomains in that direction. Therefore, the domain partitioning changes from an essentially 2D setup to a 3D configuration which requires inter node communication in $z$ direction as well. Even though the scenario was chosen carefully there are still many additional factors which might influence the overall performance. The initial refinement step introduces more subdomains which will lead to longer communication paths if they are assigned to distant processes. Also the internal data structures might show different performance characteristics for a different number of particles. To pinpoint the exact source of the deviating performance an even simpler setup together with a more detailed time measurement should be used. This will be part of future work.

\subsection{Large Size Problem}
\label{sec:LargeProblem}

\begin{figure}[h]
  \centering
  \subfloat[maximum number of particles per process]{\includegraphics[width = 0.45\textwidth]{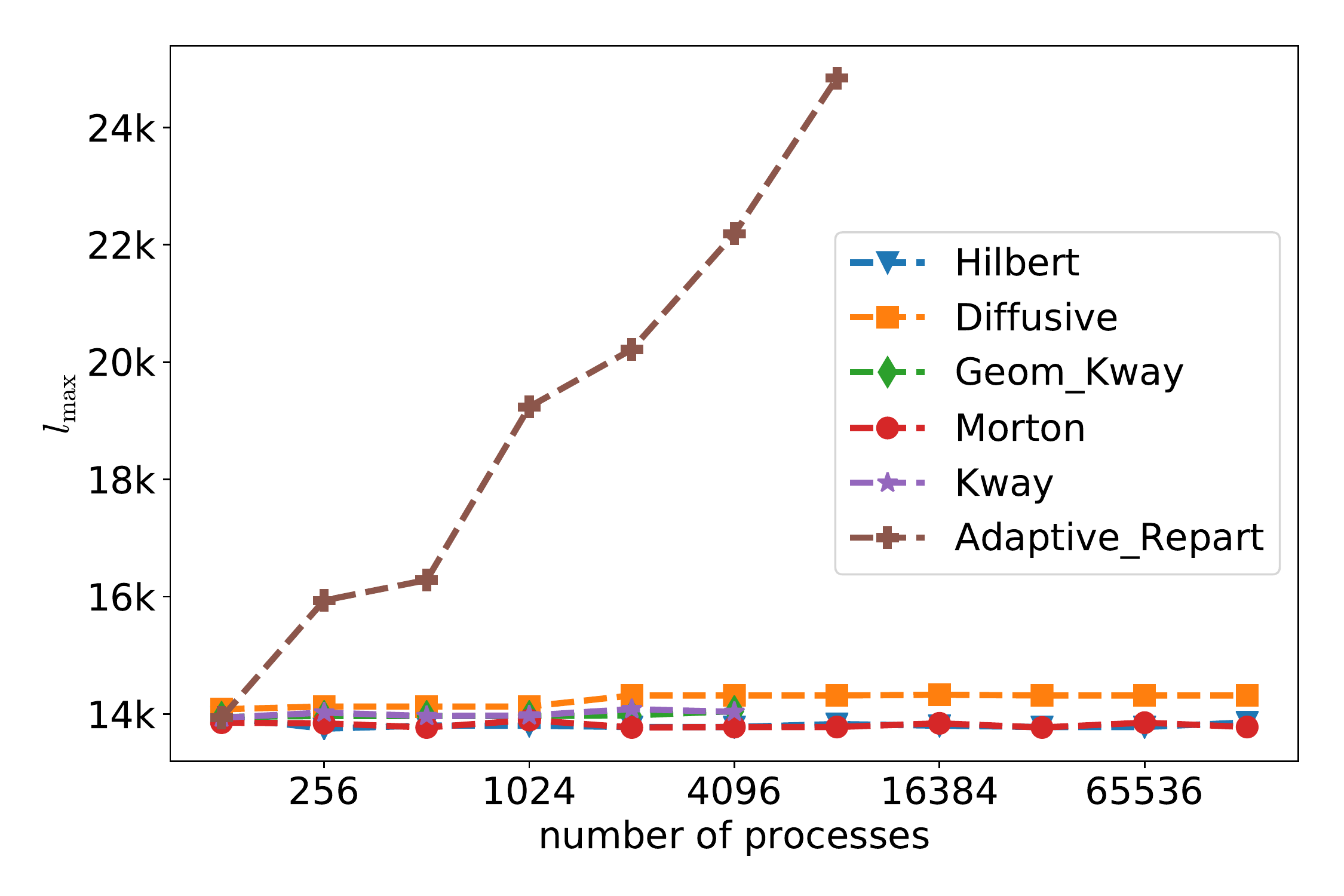}}
  \hfill
  \subfloat[relative performance after load balancing]{\includegraphics[width = 0.45\textwidth]{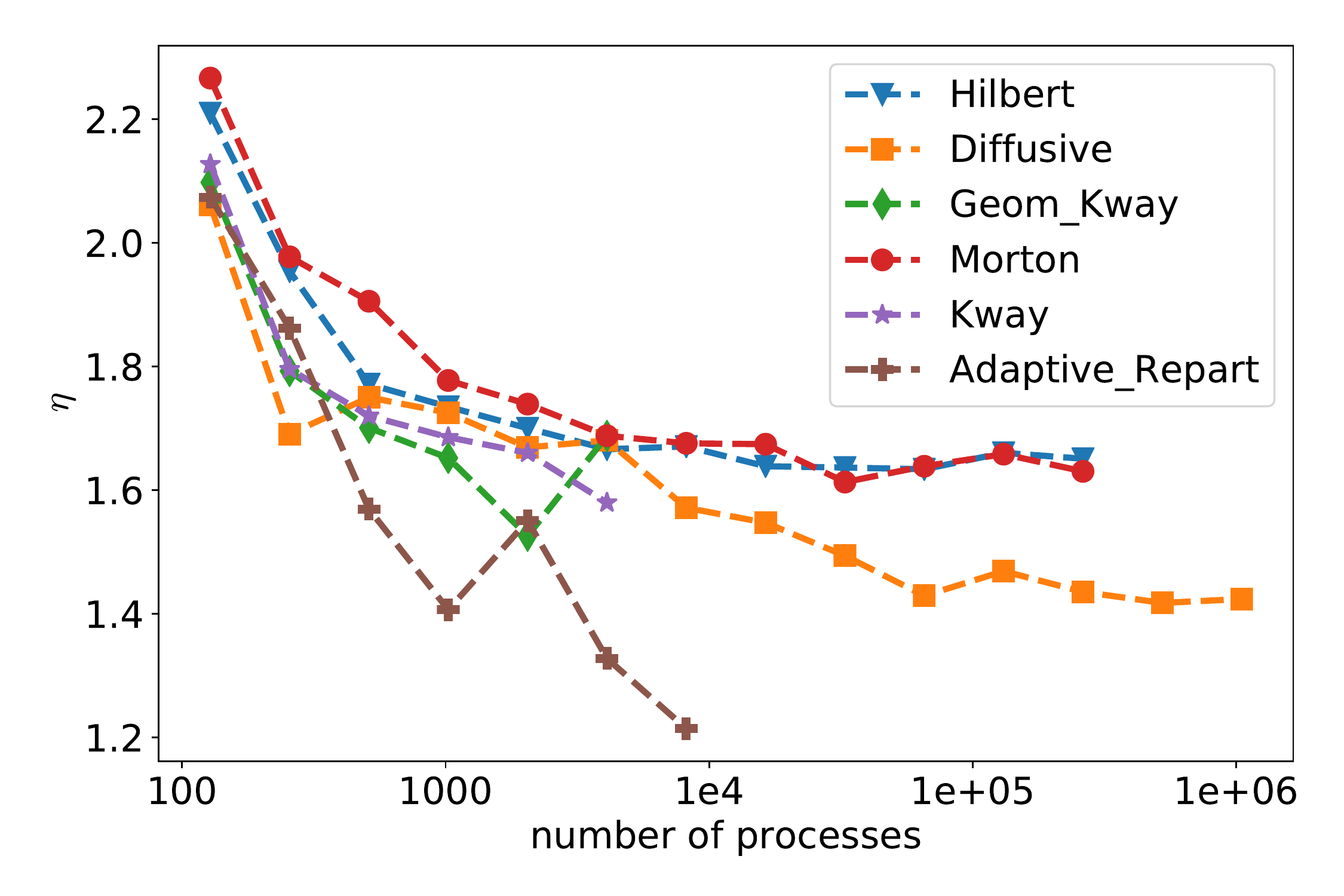}}
  \caption{On the left side the maximum load per process is shown. The right side shows a comparison of six different load balancing algorithms. Data points are collected for different simulation sizes.}
  \label{fig:LowLoad}
\end{figure}

For the second benchmark we use a half filled simulation box and less particles per subdomain. The number of particles for a completely filled subdomain is approximately \num{22000} in this setup. The smaller number of particles allows us to use four processes per core -- using full SMT capabilities -- on the Juqueen supercomputer since less memory is required per subdomain. From a performance point of view this is the optimal configuration for the rigid body dynamics simulation~\cite{Eibl2017}. Furthermore, this allows us to study the scalability up to one million processes.

In theory, a maximal performance gain of 2 can be achieved in this scenario. For this scenario we compare Hilbert and Morton space filling curves, the diffusive algorithm as well as Geom\_Kway, Kway and Adaptive\_Repart from the ParMetis library.

Fig.~\ref{fig:LowLoad}a) shows again the maximum load per process. The average workload stays the same in this scenario ($\num{11000}$). However, due to the finer resolution of the subdomains ($\num{22000} / 8 \approx \num{2750}$) the balancing quality is closer to the optimum. As can be seen in Fig.~\ref{fig:LowLoad}a) all algorithms achieve the same balancing quality, except Adaptive\_Repart. This algorithm exhibits problems getting close to the optimum when the number of processes increases. The assumption on the maximum performance gain of the computational part can now be refined also for this scenario: $\frac{\num{22000}}{\num{14000}} \approx \num{1.6}$. However, the communication will not be improved at all (twice the amount of communication and twice the amount of resources).

The performance gain is shown in Fig.~\ref{fig:LowLoad}b). All algorithms were run in all scenarios. Missing data points indicate that the simulation ran out of memory for this configuration. The same general behaviour as for the medium size problem can be observed. In the beginning, the performance gain is better than expected. It then converges towards \num{1.6} for the best performing algorithms. This indicates that the computational workload is dominating in this configuration. For small problem sizes all algorithms achieve a relatively good performance gain with the SFCs yielding the best results. For large scenarios the SFCs reach a performance gain of slightly above \num{1.6} which coincides with the predicted value. The Geom\_Kway and Kway algorithms perform slightly worse than the SFCs and they can only be used for up to \num{4096} processes. The diffusive algorithm keeps loosing performance until \num{65536} processes. Beyond this the performance stabilizes at $\approx\num{1.4}$. The diffusive algorithm is the only algorithm that can be used for all problem sizes. The Adaptive\_Repart algorithm performs worst of all algorithms, yielding only a gain of $\approx\num{1.2}$ at a maximal number of \num{8192} processes.

The general performance of the SFCs is comparable to the previous evaluation (see Sec.~\ref{sec:MediumProblem}). The same reasons apply here. Since the Geom\_Kway algorithm is partly based on SFCs, a similar performance can be expected. Unfortunately no improved performance compared to the plain SFCs can be observed with this more complicated algorithm. The plain Kway algorithm produces similar results to the Geom\_Kway algorithm. The other ParMetis algorithm, Adaptive\_Repart, shows poor performance even for moderate process numbers. All ParMetis based algorithms run out of memory quickly. The reason for this lies in the implementation of the ParMetis library. Unfortunately, an analysis of the implementation to pinpoint the actual cause is outside the scope of this article. The SFCs implemented natively in the framework run out of memory at a later point. During the implementation, special attention was paid to the memory requirements. Nevertheless, all space filling curves run out of memory at a certain process number since they rely on an allgather operation that requires $\mathcal{O}(p^2)$ memory, as explained in Sec.~\ref{sec:LoadBalancingAlgorithms}. This limits the applicability of SFCs for extreme scale parallel simulations. A better implementation might reduce the memory requirement by a constant factor, but the asymptotic complexity will still be quadratic. The diffusive algorithm has no problems with memory requirements since it is a strictly local (only stores information about and communicates only with next neighbors). However, the performance gain is worse compared to SFCs. One likely reason for this is fact that the diffusive algorithm does not maintain spatial locality. The diffusive process will often separate adjacent subdomains onto distant processors which will require communication over longer distances. Especially in Juqueen's torus topology this will reduce the performance significantly when the data must be sent across multiple nodes.

\begin{figure}[h]
  \centering
  \includegraphics[width=0.7\textwidth]{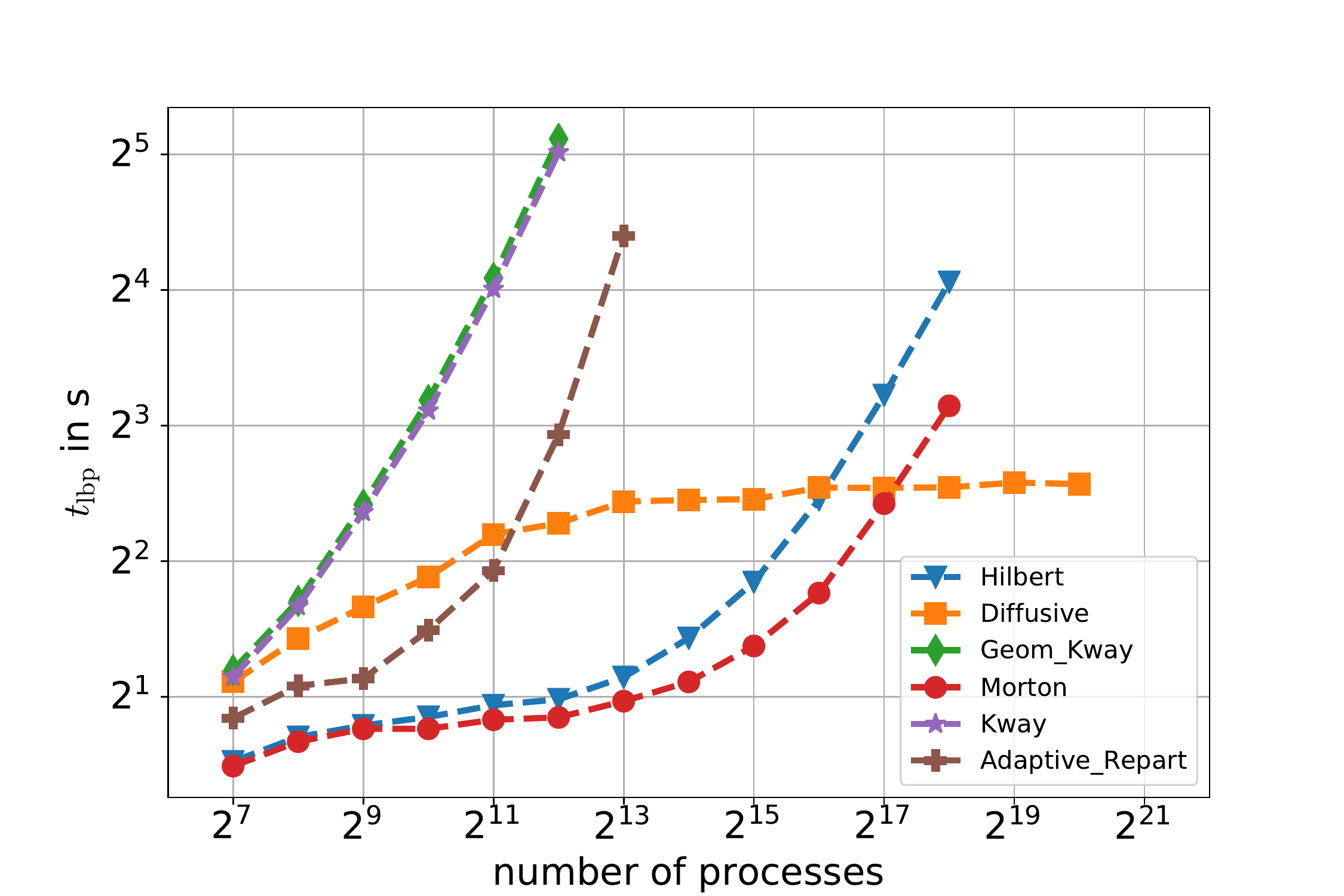}
  \caption{Runtime of various load balancing algorithms in a loglog plot. A weak scaling experiment shows the evolution of the runtime as more and more processes are used.}
  \label{fig:Runtime}
\end{figure}

Finally we compare the runtime of the load balancing algorithms. The results are displayed in Fig.~\ref{fig:Runtime}. Geom\_Kway and native Kway from the ParMetis library show quadratic runtime complexity. Although this is not significant for small process numbers, it will become a problem when progressing to extreme scale. The third algorithm of the ParMetis library Adaptive\_Repart shows linear runtime complexity but runs out of memory early. For the SFCs, a linear runtime complexity is expected since every process must compute the partitioning for all p processes. The runtime complexity of the allgather operation can be neglected as it can be implemented in $\mathcal{O}(\log p)$ complexity~\cite{Bruck1997}. The diffusive algorithm exhibits a linear increase for small process numbers, but asymptotically reaches a constant runtime complexity for large process numbers. This is also expected since it communicates only with next neighbors. In the case of the octree structure, as used in our implementation, every process has only a constant number of communication partners. The runtime of the complete load balancing pipeline for all load balancing algorithms ranges between \SI{1}{\second} and \SI{32}{\second}. Compared to a typical time step duration before load balancing of $\approx\SI{6}{\second}$ the runtime of the load balancing pipeline has to be considered and the load balancing should only be rerun when necessary. 

\section{Hopper Discharge}
\label{sec:HopperDischarge}
\begin{table}
\centering
\begin{tabular}{ l l }
  \toprule
  nodes & \num{27} \\
  cores (initial arrangement) & \num{1296} ($12\times12\times9$) \\
  particles & \num{8371596} \\
  particle shape & spheres \\
  particle radius & \num{0.5} \\
  particle distance & \num{1.0} \\
  initial particle velocity & random, between \num{-1.0} and \num{1.0} per axis \\
  particle arrangement & simple cubic \\
  cone angle & \SI{45}{\degree} \\
  orifice radius & \num{46} \\
  interaction model & hard contacts \\
  friction model & maximum dissipation~\cite{Preclik2017a} \\
  gravity & (0,0,-1) \\
  dt & \num{0.01} \\
  simulation steps & \num{20000} \\
  \bottomrule
\end{tabular}
\caption{Parameters for the hopper discharge simulation.}
\label{tab:HopperDischargeParameters}
\end{table}

In the last chapter we evaluated the performance of different load balancing algorithms in an academic scenario. This allowed us to control many potential influences on the measurements which we cannot control in real world scenarios. It also was designed in a way that only few time steps are needed to get meaningful results which saves precious compute time especially for very large runs. However, the algorithms have to perform well in applied scenarios to be of use. Therefore in a second step we evaluate the performance of the algorithms in a hopper discharge scenario. Since the ParMetis based algorithms are not suitable for large scale simulations as we have seen in the previous chapter we only evaluate the SFCs and the diffusion based algorithm against an unbalanced simulation. 

\begin{figure}[h]
  \centering
  \subfloat{\includegraphics[width = 0.3\textwidth]{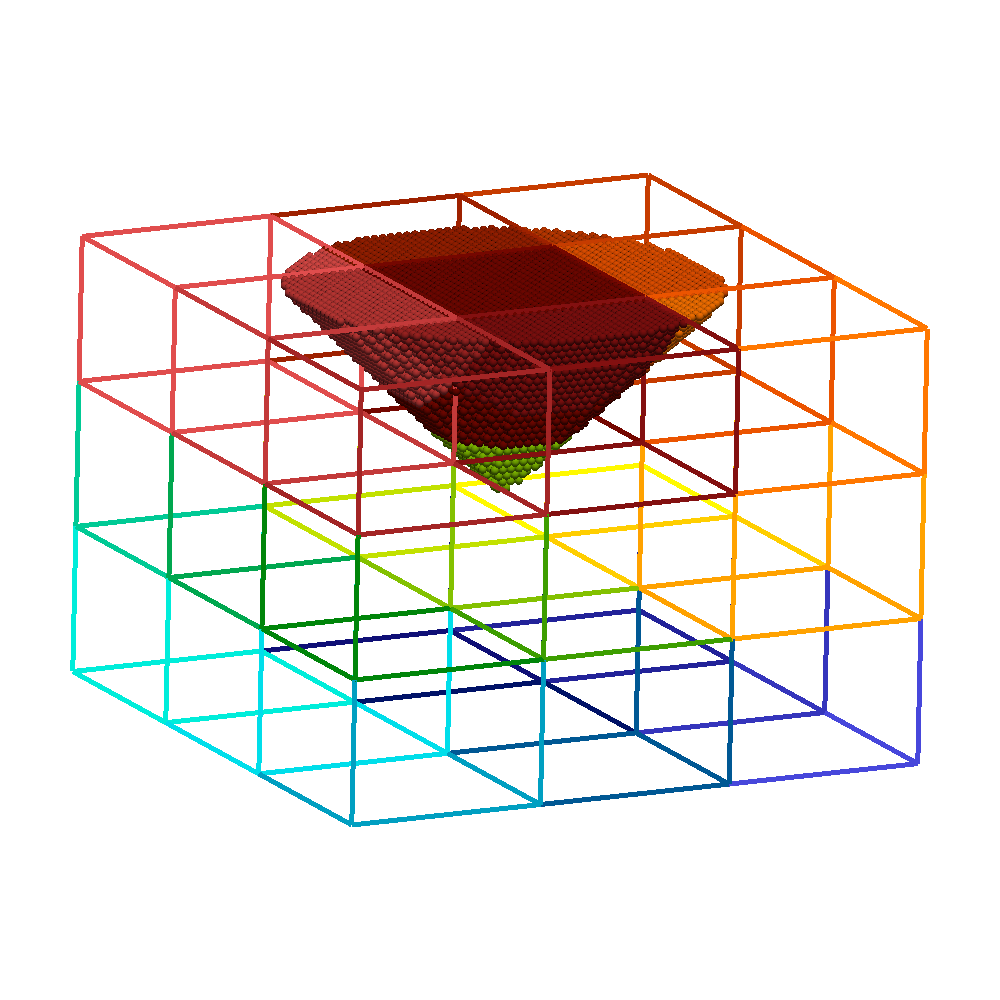}}
  \hfill
  \subfloat{\includegraphics[width = 0.3\textwidth]{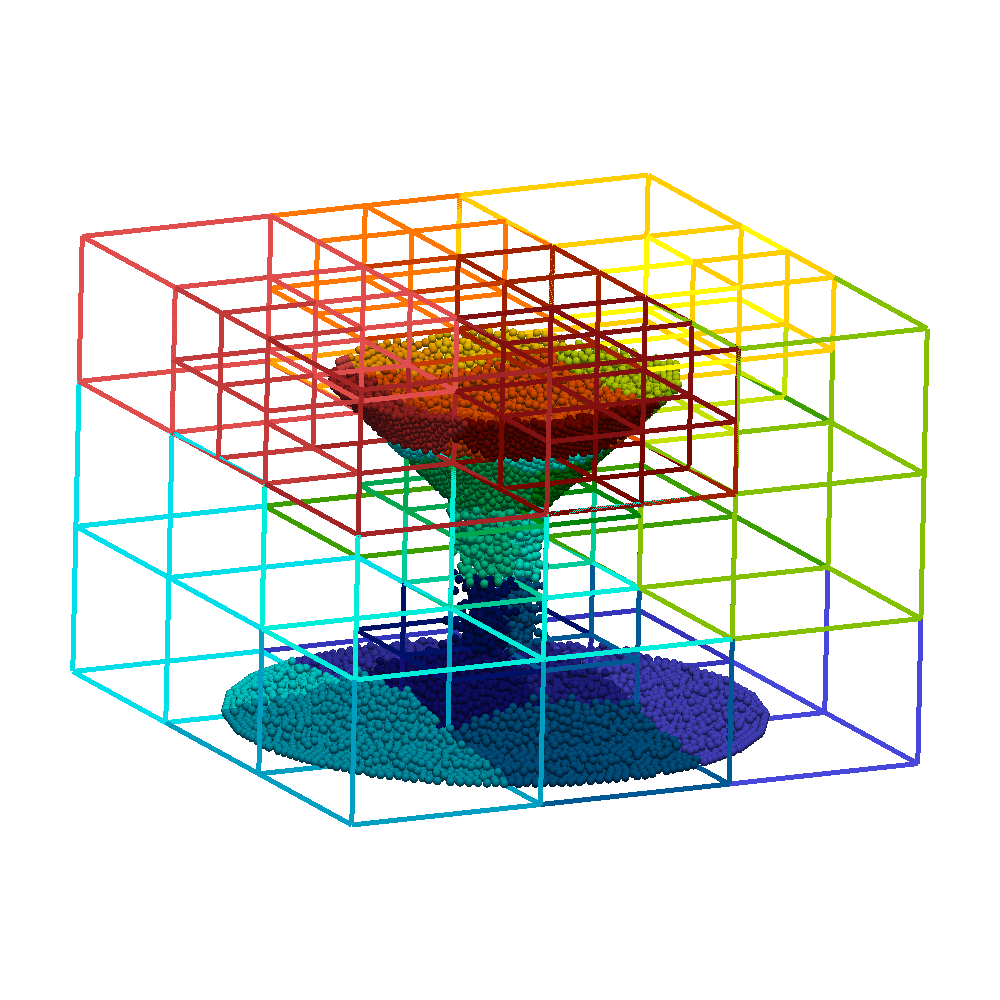}}
  \hfill
  \subfloat{\includegraphics[width = 0.3\textwidth]{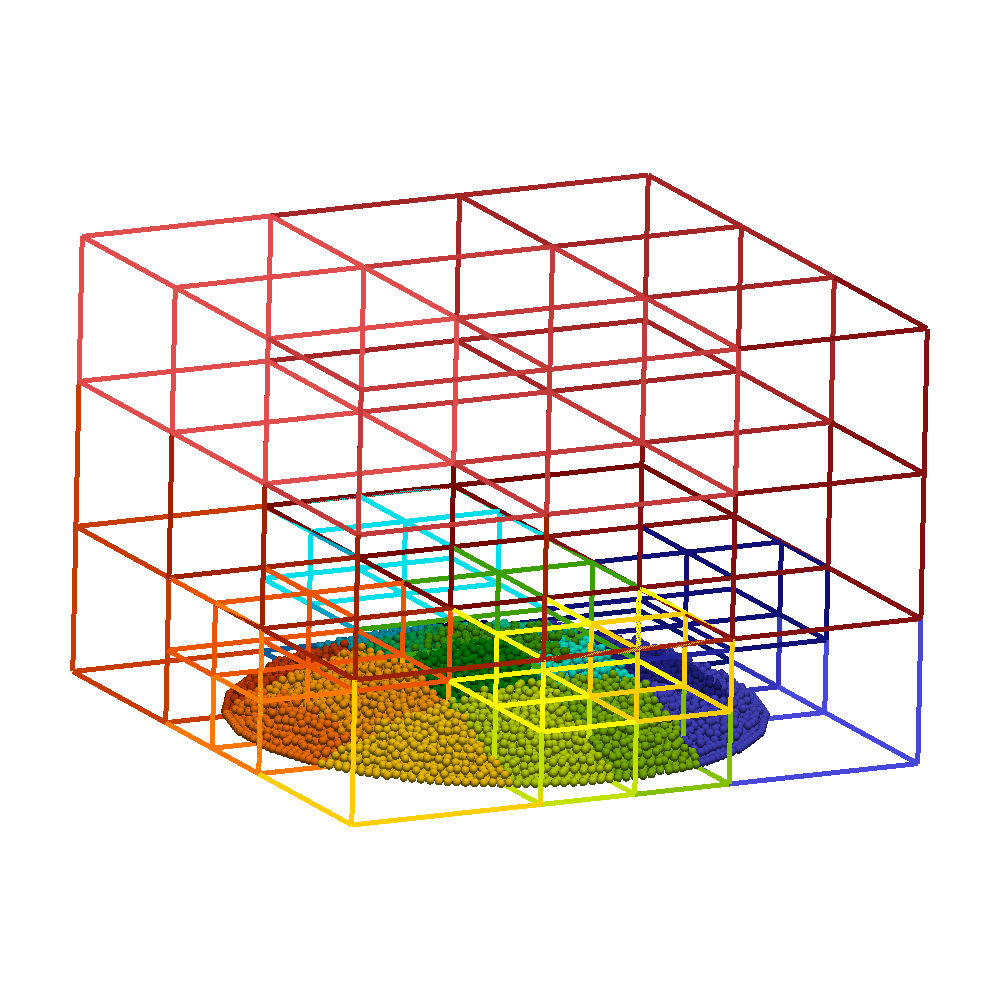}}
  \caption{From left to right: Evolution of the hopper discharge simulation in time. For a better visualization the simulation shown is much smaller then the simulation used for measurement. The picture series shows the domain partitioning and the spherical particles inside the hopper. The color coding denotes the assignment to different processes. \num{27} processes are used for this simulation. The simulation is conducted with refinement and load balancing enabled.}
  \label{fig:HopperDischarge}
\end{figure}

An illustration of the simulation is found in Fig.~\ref{fig:HopperDischarge}. The parameters for this setup are listed in Tab.~\ref{tab:HopperDischargeParameters}. Initially the simulation domain is partitioned into \num{1296} subdomains arranged in a $12\times12\times9$ lattice which are mapped 1:1 onto different cores. All initial subdomains have exactly the same size. At the beginning of the simulation all \num{8371596} particles are located inside the hopper cone and are initialized with a random velocity. The collecting tank below is empty (see Fig.~\ref{fig:HopperDischarge} left). For the static domain partitioning this situation is very imbalanced since all subdomains in the lower half part are empty. During the simulation the situation gets better as some particles are still in the hopper but some are also in the collecting tank (see Fig.~\ref{fig:HopperDischarge} middle). In the end the situation worsens again (see Fig.~\ref{fig:HopperDischarge} right) as all particles drop down into the tank. To improve the situation we apply load balancing algorithms already tested in the previous chapter also to this example. In contrast to our previous setup the computational weight is not determined directly by the number of particles. The dominating factor here is the number of contacts. Therefore, we use the number of contacts in each subdomain directly as the computational weight for this subdomain. 

The whole simulation runs for \num{20000} time steps. The load balancing pipeline is run every \num{100} time steps and statistical information which is averaged over the last 100 time steps is extracted. All simulations are carried out on the new JUWELS supercomputer, successor of the JUQUEEN supercomputer. \num{27} nodes equipped with two Intel Xeon Platinum 8168 processors with 24 cores each are used. Each node comprises \SI{96}{GB} of main memory. The nodes are connected via EDR-Infiniband in a 3 level fat-tree configuration. Unfortunately we do not have any influence on which nodes we get. Therefore the connection distance might be close or far away.

\begin{figure}[h]
  \centering
  \subfloat[Total number of contacts averaged over the last 100 time steps at a specific time step.]{\includegraphics[width = 0.45\textwidth]{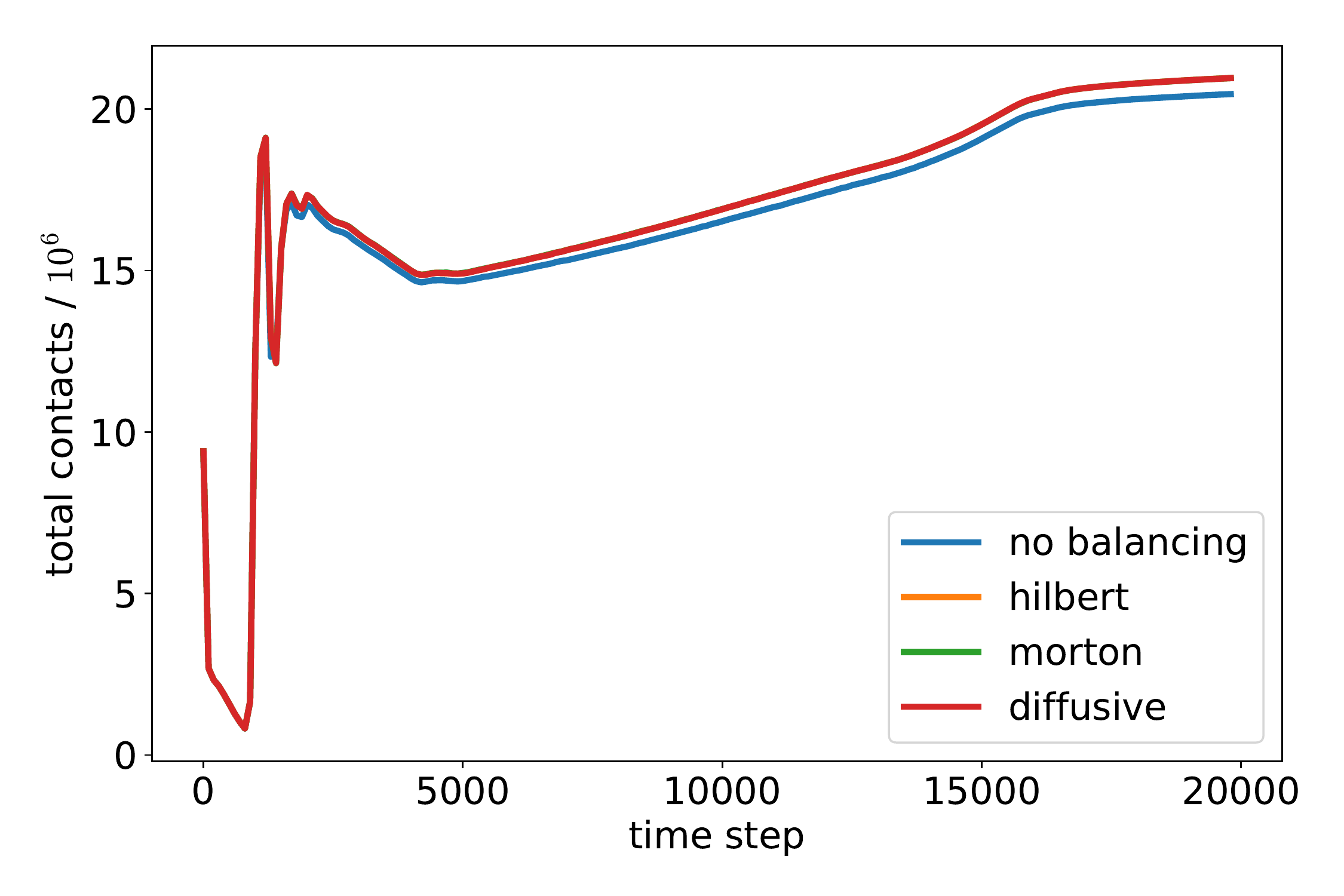}}
  \hfill
  \subfloat[Maximum number of contacts any subdomain had averaged over the last 100 time steps. The number of contacts every subdomain would have in a perfectly balanced simulation is denoted as optimum.]{\includegraphics[width = 0.45\textwidth]{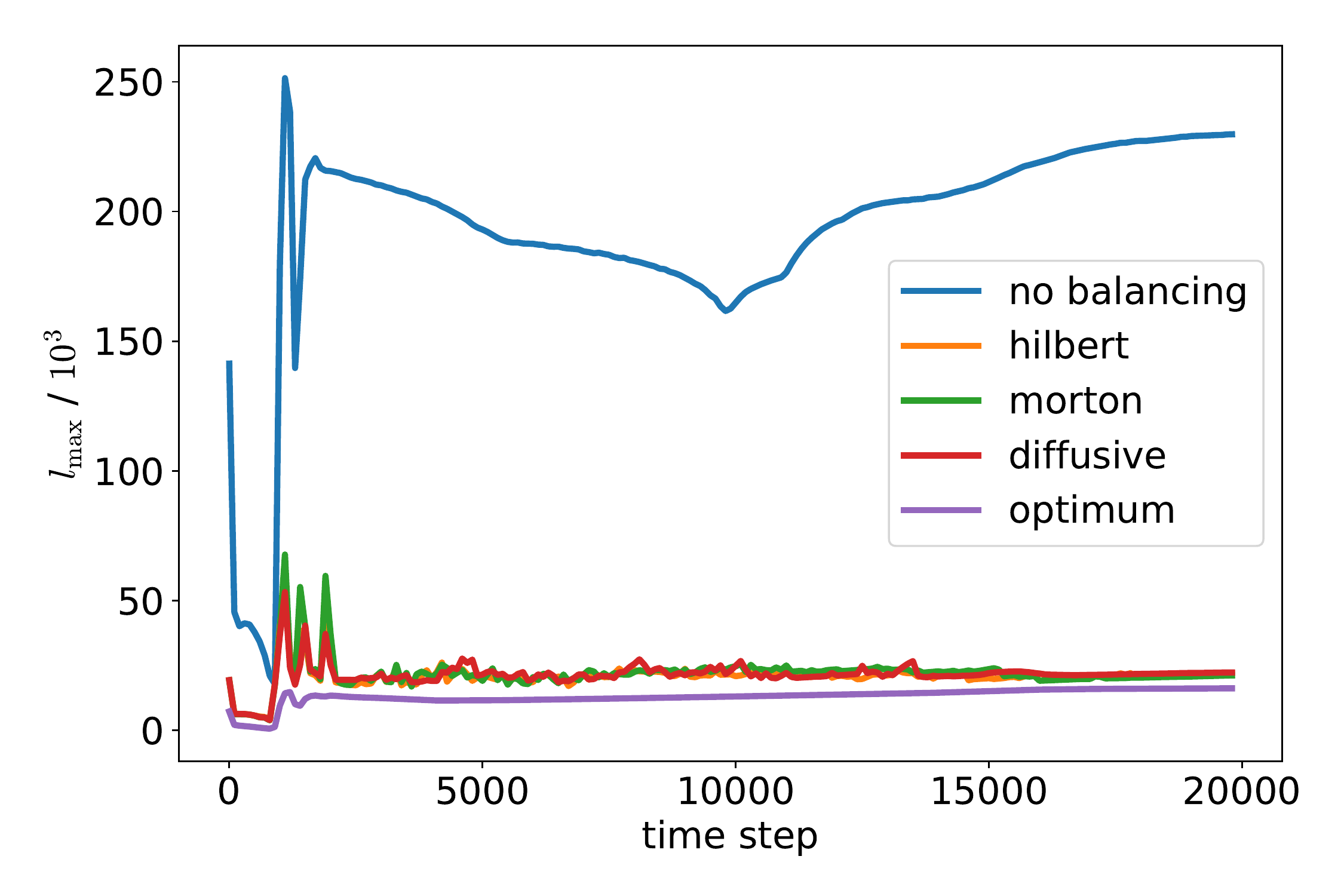}}
  \caption{Analysis of the number of contacts during a hopper discharge scenario. Simulations with and without load balancing are compared.}
  \label{fig:HopperDischargeContacts}
\end{figure}

The number of contacts throughout the simulation, as a measure for the computational weight, is shown in Fig.~\ref{fig:HopperDischargeContacts}. One can clearly see the progress of simulation in the total number of contacts for the unbalanced simulation (see Fig.~\ref{fig:HopperDischargeContacts}a)). Initially the bulk of spheres is colliding and then separating slightly explaining the very first peak. After that, the particles drop down into the hopper cone giving raise to a second spike on impact. Since some particles will bounce of once or twice the number of contacts fluctuates slightly till the particles come to a rest and start dropping down into the tank. Approximately at time step \num{4000} half of the particles have fallen down and the number of contacts reaches its minimum. After that the particles start to pile up inside the tank. A slight deviation can be seen between the unbalanced simulation and the simulations with load balancing enabled. This can be explained with the refinement process applied during the load balancing pipeline. This process introduces more and smaller subdomains increasing the interface area between subdomains which leads to an increased number of ghost particles. These additional particles are responsible for the slightly increased number of collisions.

The maximum load per process (\maxload) is shown in Fig.~\ref{fig:HopperDischargeContacts}b). Please note that this is now calculated using the number of contacts. The spikes in the beginning directly related to the changes in the total number of contacts as detailed in the previous paragraph. One can note, however, that \maxload~for the unbalanced simulation reaches its local minimum at around time step \num{10000} which is way later than the local minimum for the total number of contacts at around time step \num{4000}. So there is no direct relation between the total number of contacts and \maxload~since also the partitioning plays a huge role. After the fluctuations in the beginning \maxload~for the balanced simulations stays just above the optimum for all tested algorithms. Again most probably the optimum is not reached exactly due to the limited granularity of the domain partitioning. However, on average the balanced simulations run at only $\approx \nicefrac{1}{8}$ of the \maxload~ of the unbalanced simulation.

\begin{figure}[h]
  \centering
  \subfloat[Runtime of the last 100 time steps at a specific time step.]{\includegraphics[width = 0.45\textwidth]{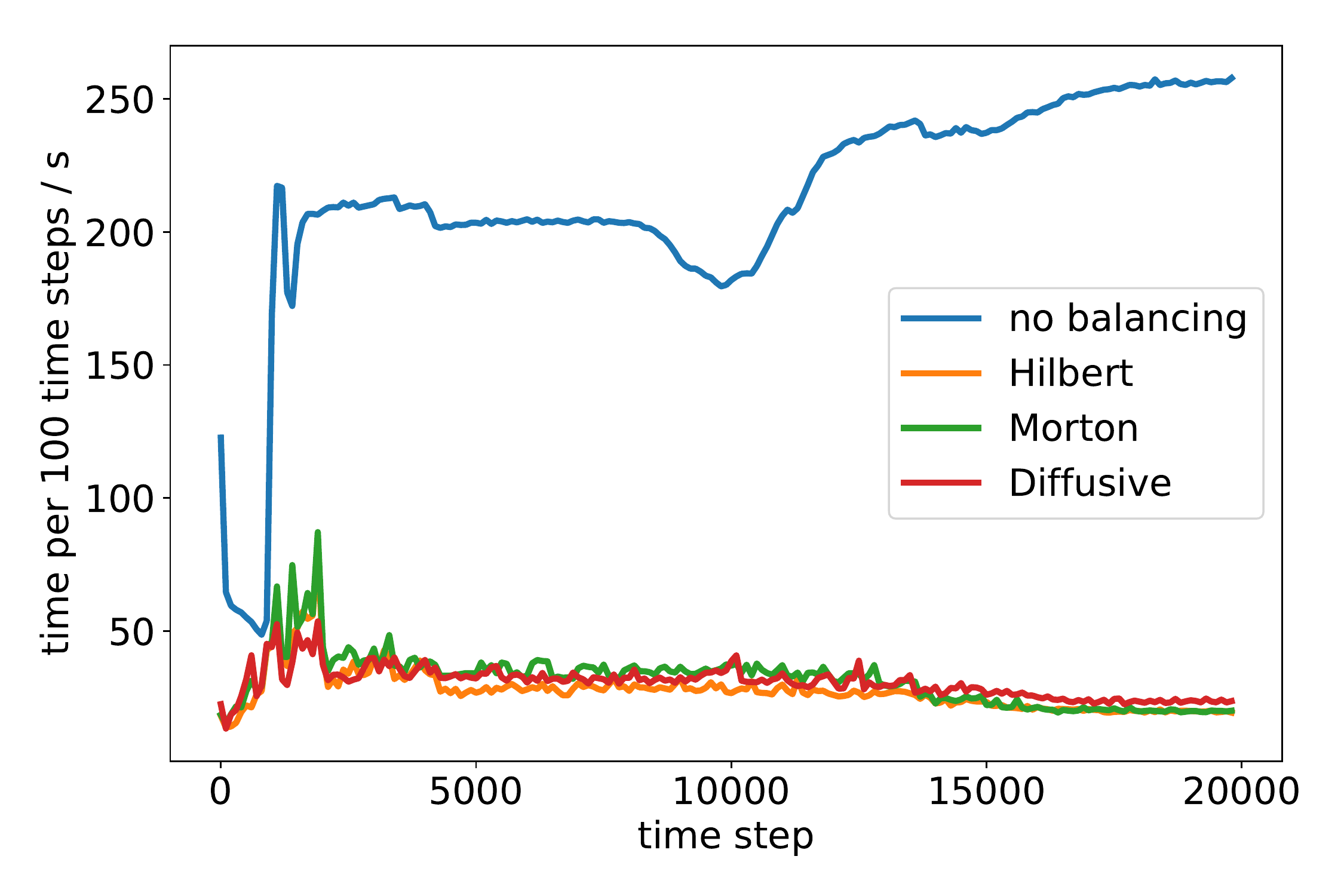}}
  \hfill
  \subfloat[Total runtime of simulation.]{\includegraphics[width = 0.45\textwidth]{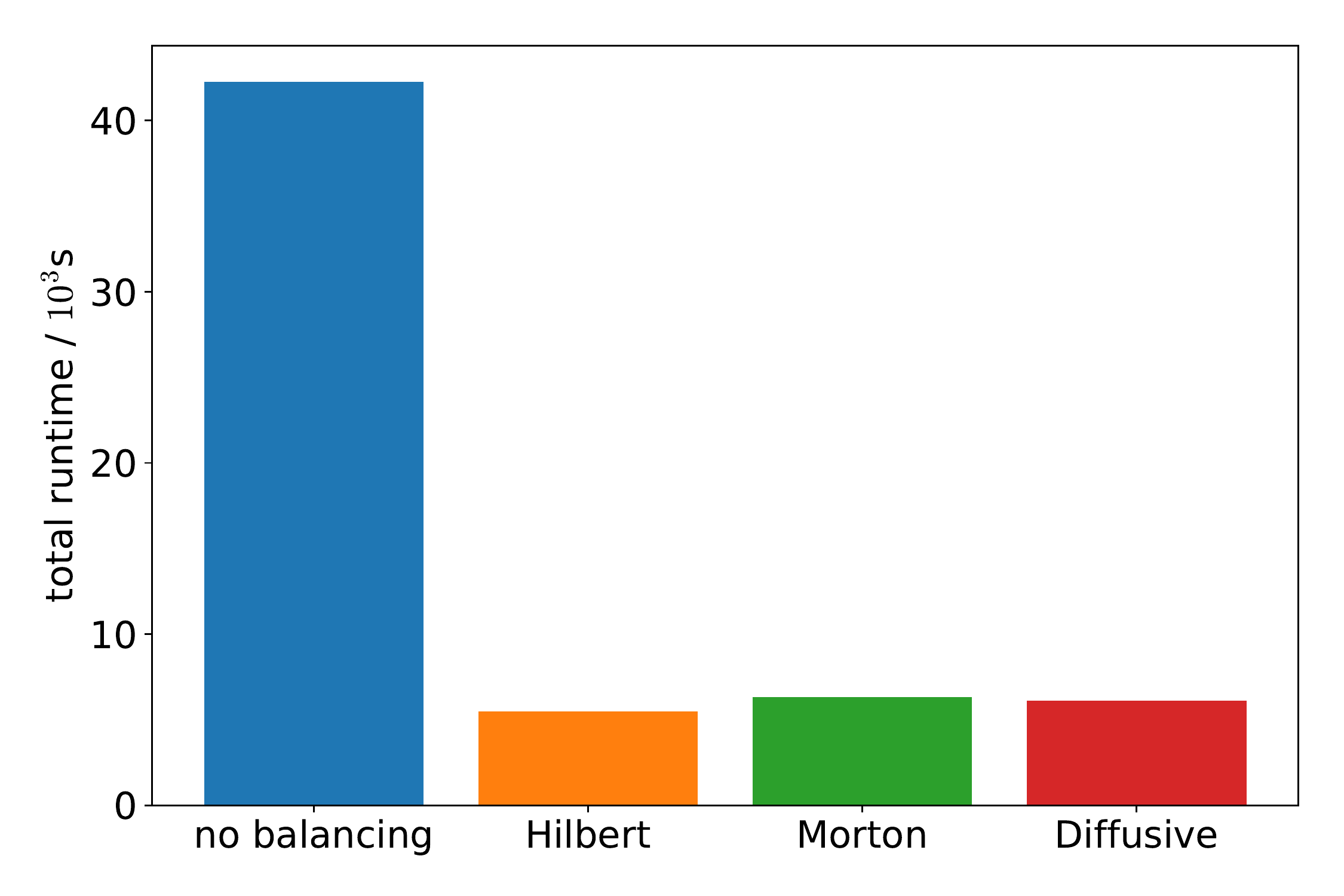}}
  \caption{Runtime comparison of a hopper discharge simulation with and without load balancing.}
  \label{fig:HopperDischargeRuntime}
\end{figure}

Since \maxload~ is a user defined quantity one also has to check the real runtime of the simulation. Measurements of the time spent for the last 100 time steps can be seen in Fig.~\ref{fig:HopperDischargeRuntime}a). The runtime relates very well to the \maxload. The local minimum of the unbalanced simulation can be seen around time step \num{10000}. A clear indication that the runtime is limited by the process with the most computational workload as it perfectly matches the minimum in Fig.~\ref{fig:HopperDischargeContacts}b). The total runtime of the whole simulation is depicted in Fig.~\ref{fig:HopperDischargeRuntime}b). The fastest simulation with load balancing enabled (Hilbert, \SI{5.5e3}{\second}) is by a factor of \num{7.6} faster than the unbalanced simulation (\SI{42e3}{\second}). This is in perfect relation to the ratio of \maxload~ between the two simulations. All simulations with load balancing enabled are very close together regardless of the algorithm used. The time needed by all load balancing pipeline executions stays below \SI{1}{\percent} of the total runtime and is therefore neglected.

\section{Conclusion}
\label{sec:Conclusion}
In this paper we have evaluated what performance gain can be expected by using different load balancing algorithms in rigid particle dynamics simulations. We have used a carefully selected simulation setup that allowed us to give a-priori estimates on the expected performance gain. Different load balancing algorithms, namely graph based algorithms from the ParMetis library, a diffusion based algorithm, and space filling curves, have been evaluated and compared to our predictions. We have also studied the resources required by these algorithms with respect to memory and runtime. All these experiments were executed with different problem sizes to evaluate the suitability of these algorithms for extreme scale parallel environments. \correct{}{The real world applicability of the load balancing algorithms was subsequently analyzed in a hopper discharge scenario.}

\correct{O}{For the artificial scenario o}nly the diffusive algorithm was usable up to the maximum number of processes used for this evaluation. All other algorithms exhausted the available memory earlier at a smaller number of processes. Also with respect to runtime, the diffusive algorithm was the only one showing a constant runtime complexity. With respect to this criterium, the SFCs performed second best, having a linear runtime complexity. The Kway algorithm with quadratic complexity came in last.

However, the performance gain achieved by the diffusive algorithm is worse compared to SFC based algorithms. This leads to the conclusion that for small to medium size parallel applications, SFCs are currently the best choice. At very large scale, currently only the diffusive algorithm is applicable, however, its performance is only suboptimal. It is also clearly shown that the maximal performance gain is limited by the forest of octrees partitioning regardless of the balancing algorithm used. Since a perfect equal load distribution is highly unlikely to be reached by all processes due to the granularity of the load balancing, the maximal performance gain will always be less than the theoretical optimal value.

The measurements shown in this paper also indicate that \correct{}{even though we tried to minimize potential side effects} there are \correct{}{still} other factors influencing the load balancing which were not considered here. Future investigations should pinpoint the exact cause for performance deviations of all algorithms to identify areas for improvement. The diffusive algorithm is already a promising candidate for load balancing in extremely parallel environments. However, the performance gain by diffusive balancing must be improved to become competitive with SFCs for medium size problems.

\correct{}{Finally the hopper discharge simulations show that SFC based algorithms as well as diffusive ones can be successfully applied to small scale problems and achieve substantial performance improvements. This additional performance gained by using load balancing can drastically reduce the computation time.}

\correct{}{The analysis presented in this paper assumes that the workload is distributed onto equal worker units. It is therefore also applicable to simulations making use of accelerators if the applied framework supports that. However, in its current state it cannot be applied to heterogeneous worker units.}

\section*{Acknowledgment}
The authors gratefully acknowledge the Gauss Centre for Supercomputing e.V. (www.gauss-centre.eu) for funding this project by providing computing time through the John von Neumann Institute for Computing (NIC) on the GCS Supercomputers JUQUEEN and JUWELS at Jülich Supercomputing Centre (JSC).

The authors would like to acknowledge the support through the Cluster of Excellence Engineering of Advanced Materials (EAM).

Declaration of interest: none

\section*{References}

\bibliography{literature_clean}

\end{document}